\documentclass[numbers]{elsarticle}
\usepackage{latexsym}
\usepackage{natbib}
\usepackage[latin2]{inputenc}
\usepackage[hidelinks]{hyperref}
\usepackage{longtable, lineno}

\makeatletter
\def\ps@pprintTitle{%
	\let\@oddhead\@empty
	\let\@evenhead\@empty
	\def\@oddfoot{\centerline{\thepage}}%
	\let\@evenfoot\@oddfoot}
\makeatother

\begin{document}

\title{How to Direct the Edges of the Connectomes: Dynamics of the Consensus Connectomes and the Development of the Connections in the Human Brain}


\author[p]{Csaba Kerepesi}
\ead{kerepesi@pitgroup.org}
\author[p]{Balázs Szalkai}
\ead{szalkai@pitgroup.org}
\author[p]{Bálint Varga}
\ead{balorkany@pitgroup.org}
\author[p,u]{Vince Grolmusz\corref{cor1}}
\ead{grolmusz@pitgroup.org}
\cortext[cor1]{Corresponding author}
\address[p]{PIT Bioinformatics Group, Eötvös University, H-1117 Budapest, Hungary}
\address[u]{Uratim Ltd., H-1118 Budapest, Hungary}

\date{}


\begin{abstract}
The human braingraph or the connectome is the object of an intensive research today. The advantage of the graph-approach to brain science is that the rich structures, algorithms and definitions of graph theory can be applied to the anatomical networks of the connections of the human brain. In these graphs, the vertices correspond to the small (1-1.5 cm$^2$) areas of the gray matter, and two vertices are connected by an edge, if a diffusion-MRI based workflow finds fibers of axons, running between those small gray matter areas in the white matter of the brain. 
One main question of the field today is discovering the directions of the connections between the small gray matter areas.
In a previous work we have reported the construction of the Budapest Reference Connectome Server \url{http://connectome.pitgroup.org} from the data recorded in the Human Connectome Project of the NIH.The server generates the consensus braingraph of 96 subjects in Version 2, and of 418 subjects in Version 3, according to selectable parameters. After the Budapest Reference Connectome Server had been published, we recognized a surprising and unforeseen property of the server. The server can generate the braingraph of connections that are present in at least $k$ graphs out of the 418, for any value of $k=1,2,...,418$. When the value of $k$ is changed from $k=418$ through $1$ by moving a slider at the webserver from right to left, certainly more and more edges appear in the consensus graph. The astonishing observation is that the appearance of the new edges is not random: it is similar to a growing tree. We refer to this phenomenon as the dynamics of the consensus connectomes. We hypothesize that this movement of the slider in the webserver may copy the development of the connections in the human brain in the following sense: the connections that are present in all subjects are the oldest ones, and those that are present only in a decreasing fraction of the subjects are gradually the newer connections in the individual brain development. An animation on the phenomenon is available at \url{https://youtu.be/EnWwIf_HNjw}. Based on this observation and the related hypothesis, we can assign directions to the edges of the connectome as follows: Let $G_{k+1}$ denote the consensus connectome where each edge is present in at least $k+1$ graphs, and let $G_k$ denote the consensus connectome where each edge is present in at least $k$ graphs. Suppose that vertex $v$ is not connected to any other vertices in $G_{k+1}$, and becomes connected to a vertex $u$ in $G_k$, where $u$ was connected to other vertices already in $G_{k+1}$. Then we direct this $(v,u)$ edge from $v$ to $u$.

\end{abstract}

\maketitle

\section{Introduction} 

 The Human Connectome Project \cite{McNab2013} has produced high-quality MRI-imaging data of hundreds of healthy subjects. The enormous quantity of data is almost impossible to use in brain research without introducing some rich structure that helps us to get rid of the unimportant details and allow us to focus on the essential data in the set. We believe that the braingraph or the connectome is such a structure to apply.

 The braingraphs or connectomes are discretizations of the diffusion MRI imaging data. Being a graph, it has a set of vertices and some pairs of these vertices are the edges of the graph. Each vertex corresponds to a small (1-1.5 cm$^2$) areas (called Regions of Interest, ROIs) of the gray matter, and two vertices are connected by an edge, if a diffusion-MRI based workflow finds fibers of axons, running between those ROIs in the white matter of the brain. In other words, the braingraph concentrates on the connections between areas of gray matter (this is an essential part of the data) and forgets about the exact spatial orbits of the axon-fibers, running between these gray matter areas in the white matter of the brain (these are the unimportant part of the data). The braingraphs may record the length or the width of these fibers as edge-weights but definitely does not contain any spatial description of their orbit in the white matter.
 
 An important question is the determination of the direction of the graph -- or connectome -- edges in these braingraphs. By our knowledge, the present diffusion-MRI based workflows have no data showing the direction of the neuronal fiber tracts between the ROIs. 
 
 Hundreds of publications deal with the properties of the human connectome every year (e.g., \cite{Ingalhalikar2014b,Szalkai2015,Hagmann2012,Craddock2013a}, but very few analyze the common edges and the edge-distributions between distinct subjects and distinct brain areas \cite{Szalkai2015a,Kerepesi2015c}. In \cite{Kerepesi2015c} we have mapped the inter-individual variability of the braingraphs in different brain regions, and we have found that the measure of the variability significantly differs between the regions: there are more and less conservative areas of the brain.
  
 \begin{figure}[h!]
 	\centering
 	\includegraphics[width=5in]{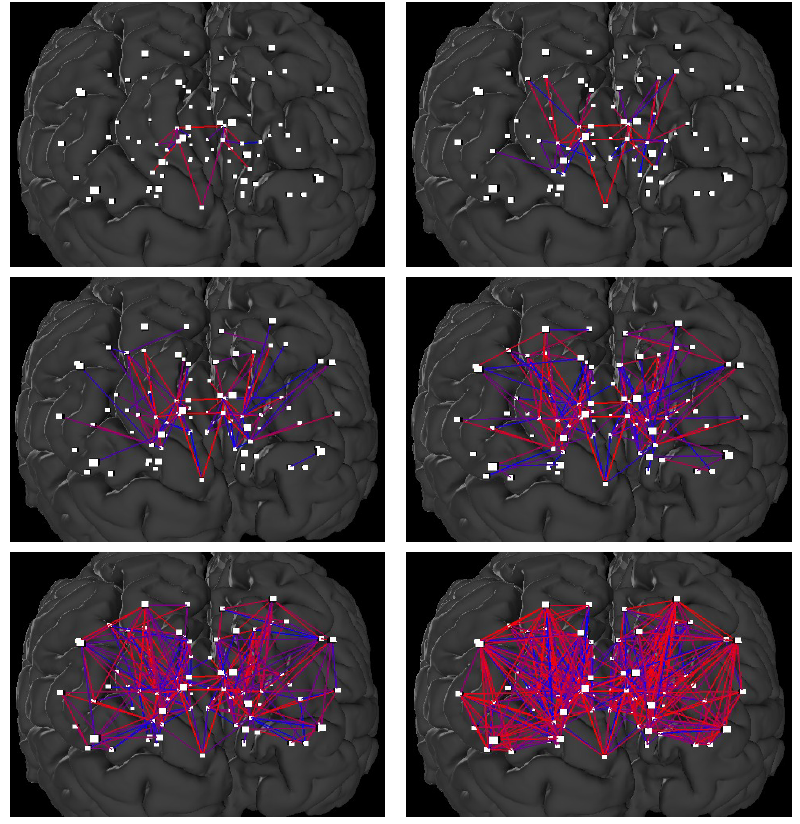}
 	\caption{Snapshots on the tree-like structure of the Budapest Reference Connectome Server v2.0. The edges of the smallest graph can be identified easily with using the webserver. For example, the edges that are present in all braingraphs include edges between Right-Caudate and Right-Pallidium, Left-Thalamus-Proper and Brain-Stem, Right-Thalamus Proper and Right-Putamen.}
 \end{figure}

 \section{Results}
 
 In the construction of the Budapest Reference Connectome Server \url{http://connectome.pitgroup.org} \cite{Szalkai2015a} not those edges were mapped that differ \cite{Kerepesi2015c}, but, on the contrary, those that are the same in at least $k$ subject's braingraphs, for $k=1,2,...,418$. These parametrized consensus-graphs describe the common connectomes of healthy humans, parametrized with $k$.
 
 For $k=418$ we get only those edges that are present in all the 418 braingraphs. For $k=1$ we get those edges that are present in at least one braingraph from these 418. Therefore, if we change the value of $k$, one-by-one, from $k=418$ through $k=1$, we will have more and more edges in the graph (Figure 1). 
  
 We have observed that the order of the appearance of the new edges when we were decreasing the value of $k$ from 418 through 1, is not random at all. More precisely, it resembles a growing tree: the newly appearing edges are usually connected to the already existing edges. This phenomenon is observable in the animation at \url{https://youtu.be/EnWwIf_HNjw} (we remark that graph-theoretically, the growing structure is not a tree as a graph). The same observation was done in Version 2 (with 96 braingraphs) and Version 3 (with 418, 476 and 477 braingraphs, depending on the fiber-numbers selected) of the server. 
 
 In what follows we clarify the implications of this observation to the 
 \begin{itemize}
 	\item[(i)] description of the individual development of the connections in the human brain, and 
 	\item[(ii)] the determination of the direction of the edges in the human connectome.
 \end{itemize}

 \begin{figure} [h!]
 	\centering
 	\includegraphics[width=5in]{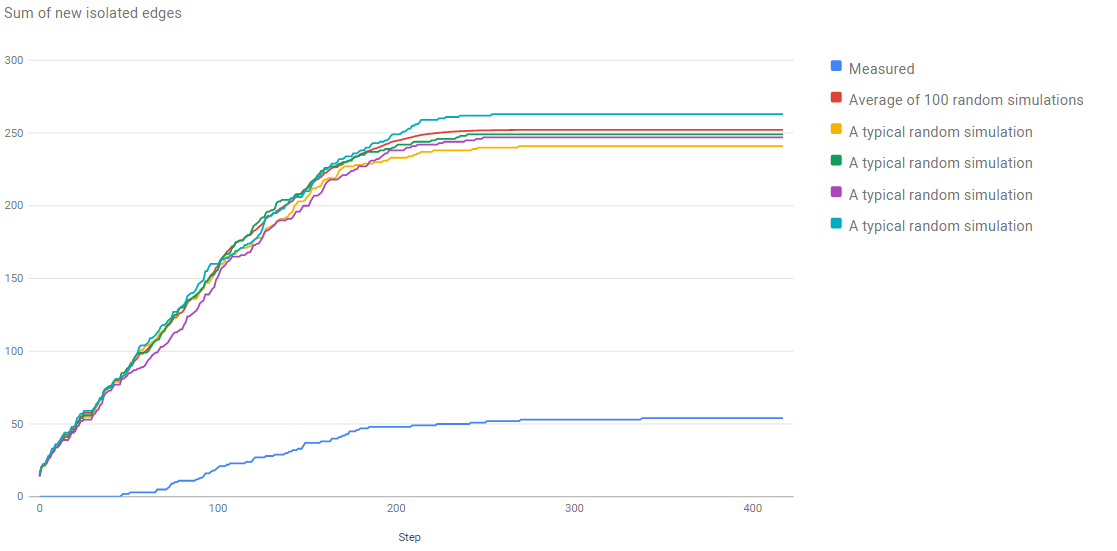}
 	\caption{The comparison of the random simulation and the real buildup of the edges in the Budapest Reference Connectome server v3.0.}
 \end{figure}
 
 \section{Discussion}
 
 The observation is verified by Figure 2, made for the Version 3.0 of the server, with 418 braingraphs. For steps $\ell=0$ through $\ell=417$, for $k=418-\ell$, we have visualized the number of those new edges (that were present in $k$ connectomes, but were not present in $k+1$ connectomes), which connect two vertices, which were not adjacent to any edges before (i.e., they were isolated vertices). We have compared 
 \begin{itemize} 
 	\item a random model, where exactly that many new edges were added randomly in uniform distribution, as in the graph generated by the Budapest Reference Connectome Server, 
 	\item and the graph of edges drawn by the Budapest Reference Connectome Server.
 \end{itemize}
 
 In the random model, in each step, the same number of edges were added to the graph randomly (independently, in uniform distribution), as in the Budapest reference Connectome Server.

 The difference is very clear on Figure 2: in the random model, dramatically more new edges appear that are not connected to the old ones. 
 
 Another visualization of this surprising phenomenon is the component tree of the evolving graph, made for Version 2 with 96 braingraphs. As $k$ decreases from 96 to 1, zero or more new edges are added to the existing graph in each step. In the step corresponding to $k$, those edges appear that are present in exactly $k$ graphs. This may result in the forming of new connected components, and/or the merging of some older components of the graph. The phenomenon can be visualized on a graph-theoretical tree, where each level of the tree corresponds to some value of $k$. On each level, some leaf nodes may appear (for each new component), and the existing nodes may merge into a parent node. We can also assign colors to the nodes according to the following scheme: the leaves get a new color, and a parent node gets the color of its child node corresponding to the largest merged component. The component-tree of the graph is visualized on a very large, labeled interactive figure  at the site \url{http://pitgroup.org/static/graphmlviewer/index.html?src=connectome_dynamics_component_tree.graphml}.
 
 We hypothesize that those edges that are contained in many of the graphs were developed in an earlier stage of the brain development than those that are present in fewer subjects. As a possible explanation, we think that those neurons that connect to the developing braingraph at \url{https://youtu.be/EnWwIf_HNjw} will not receive apoptosis signals \cite{Roth2001,Nonomura2013,gordon1995apoptosis} and will survive, while other neurons, which are not connected to the older graph, will be eliminated by receiving apoptosis signals in the individual brain development. 
 
 In other words, we assume that the connections that are present in almost all braingraphs (c.f., upper left panel of Fig.1) were developed first. Next, new connections were developed, but those neurons whose connections were disconnected from these oldest neurons were eliminated. Next, new neuronal connections were developed, but only those neurons survived that were connected to the building network. Since the deviation between the new edges among the subjects was increased step-by-step, the newer the connections, the fewer the subjects have those edges. 
 
 This assumption explains our findings, and it is in line with the ``competition hypothesis'' of the brain development \cite{gordon1995apoptosis}.
 
 \subsection{How to direct the edges of the human connectome?} 
  
For any neuron, there exists a well-defined direction of the signal propagation from the soma through its axon. Diffusion MRI-based methods can be used to identify the spatial location of the fiber tracts, consisted of axons, but their directions, by our present knowledge, cannot be discovered from the MRI data.
  
If the order of development of the edges in the connectome is known then we can easily assign a direction to those edges that connects a vertex to another one, such that the first vertex was not connected to any other vertex before, but the second vertex was already connected to the network, when we consider the transition of the edges that were present in at least $k+1$ graphs through the edges that were present in at least $k$ graphs.  
  
More exactly, the observation described above implies a straightforward method for directing some (but not all) the edges of the connectome. Consider the undirected edge ${u,v}$, and our goal is to assign a direction to this edge. Let $G_{k+1}$ denote the consensus connectome where each edge is present in at least $k+1$ graphs, and let $G_k$ denote the consensus connectome where each edge is present in at least $k$ graphs. Both $G_{k+1}$ and $G_k$ have the same set of vertices, all the edges of $G_{k+1}$ are also the edges of $G_k$, but $G_k$ typically has more edges than $G_{k+1}$. Suppose that vertex $v$ was not connected to any other vertices in $G_{k+1}$, and becomes connected to a vertex $u$ in $G_k$, where $u$ was connected to other vertices in $G_{k+1}$. Then we direct this $(v,u)$ edge from $v$ to $u$, and denote it as an ordered pair $(v,u)$ (Figure 3). Obviously, if our hypothesis is correct, then the undirected edge ${u,v}$ remained in the consensus connectome since vertex $v$ did not get an apoptosis signal, since $u$ was already been connected to the growing network.

\begin{figure} [h!]
	\centering
	\includegraphics[width=5in]{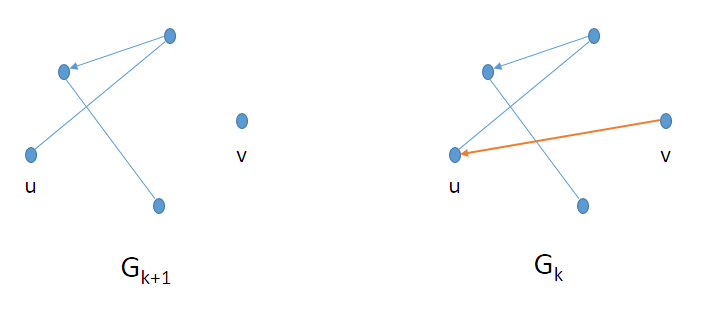}
	\caption{Let $G_{k+1}$ denote the consensus connectome where each edge is present in at least $k+1$ graphs, and let $G_k$ denote the consensus connectome where each edge is present in at least $k$ graphs. Both $G_{k+1}$ and $G_k$ have the same set of vertices, all the edges of $G_{k+1}$ are also the edges of $G_k$, but $G_k$ typically has more edges than $G_{k+1}$. The $(v,u)$ edge is directed from $v$ to $u$, if $v$ is not connected to any other vertices in $G_{k+1}$, and becomes connected to a vertex $u$ in $G_k$, where $u$ was connected to other vertices in $G_{k+1}$. Then we direct this $(v,u)$ edge from $v$ to $u$.}
\end{figure}

We remark that those new edges that connect two, previously isolated points ("isolated edges"), or those that connect two vertices, where both of them were connected to the network before, cannot be directed this way.

\section{Methods} 

The description of the program and the methods applied in the construction of the Budapest Reference Connectome Server \url{http://connectome.pitgroup.org} is given in \cite{Szalkai2015a}. 

The animation at \url{https://youtu.be/EnWwIf_HNjw} were prepared by our own Python program from the tables generated by the Budapest Reference Connectome Server \citep{Szalkai2015a} with the following settings: Version 2 (i.e., 96 subjects), Population: All (i.e., both male and female subjects), Minimum edge confidence running from 100 \% through 26\%, Minimum edge weight is 0, Weight calculation model: Median. It contains the common edges found in $k$ subject's braingraphs, from $k=96$ through $k=25$. The number of vertices is 1015.

\section{Conclusions:} We have observed that the buildup of the consensus graphs in the Budapest Reference Connectome Server is far from random when the $k$ parameter is changed from $k=418$ through 1. This observation suggests an underlying structure in the consensus braingraphs: the edges, which are present in more subjects are most probably older in the individual brain development than the edges, which are present fewer individuals. This assumption is in line with the ``competition hypothesis'' of the brain development \cite{gordon1995apoptosis}. We believe that this observation is applicable to discover the finer structure of the development of the connections in the human brain.

Based on this hypothesis we were able to assign directions to some of the otherwise undirected edges of the connectome, built through a diffusion MRI based workflow.

\section*{Data availability:} The unprocessed and pre-processed MRI data that served as a source of our work are available at the Human Connectome Project's website:

\url{http://www.humanconnectome.org/documentation/S500} \cite{McNab2013}. 

\noindent The assembled graphs that were used to build the Budapest Reference Connectome Server can be downloaded at the site \url{http://braingraph.org/download-pit-group-connectomes/}.



\section*{Acknowledgments}
Data were provided in part by the Human Connectome Project, WU-Minn Consortium (Principal Investigators: David Van Essen and Kamil Ugurbil; 1U54MH091657) funded by the 16 NIH Institutes and Centers that support the NIH Blueprint for Neuroscience Research; and by the McDonnell Center for Systems Neuroscience at Washington University.



\end{document}